\documentstyle[12pt]{article}
\begin{document}
\begin{titlepage}
\rightline{\vbox{\halign{&#\hfil\cr
&CUMQ/HEP 92\cr
&UCTP-115/96\cr
&\today\cr}}}
\vspace{0.5in}
\begin{center}
{\Large\bf Lepton-Flavor Violating Decays of the $Z$-boson in a Left-Right
Supersymmetric Model}
\\
\medskip
\vskip0.5in

\normalsize {{\bf M. Frank}$^{\rm a}$ and {\bf H. Hamidian}$^{\rm b}$}
\smallskip
\medskip

{ \sl $^a$Department of Physics, Concordia University, 1455 De Maisonneuve Blvd. W.\\ Montreal, Quebec, Canada, H3G 1M8\\ $^b$Department of Physics, University of Cincinnati\\ 
Cincinnati, Ohio, 45221 U.S.A.}\smallskip
\end{center}
\vskip1.0in

\begin{abstract}
We study the one-loop supersymmetric contributions to  
lepton-flavor violating $Z$ -boson decays in a fully left-right 
symmetric model. In addition to right-handed scalar neutrinos,
the decays could receive important contributions from doubly-charged 
triplet Higgsinos which couple to leptons only. We find that the decay 
widths reach present experimental bounds and discuss mass constraints for
the contributing supersymmetric partners.
\end{abstract}
 
\end{titlepage}
\baselineskip=20pt

\newpage
\pagenumbering{arabic}
\section{\bf Introduction}

The quest for understanding beyond the Standard Model physics and 
its implications for the present and future colliders has relied heavily on extended gauge structures of the Standard Model.
Among these Supersymmetric Grand Unfied
Theories $(SUSY GUT's)$ such as $SO(10)$ and $SU(5)$ have received significant 
attention, 
due to their correct prediction of the electroweak mixing angle~\cite{ellis}. 
These 
theories provide unfication of interactions at a higher scale than the ordinary
$GUT's$, which prevents rapid proton decay. Another attractive feature of these
models is the prediction of a small but non-vanishing neutrino mass. Indeed,  
anomalies measured in solar and atmospheric neutrino fluxes seem to lend 
support to the idea that neutrinos should have a small but nonvanishing 
mass. The most 
promising candidate for a mechanism to give the neutrino a mass is the see-saw
model, which explains the smallness of the neutrino mass in terms 
of the large Majorana mass for a right-handed neutrino $\nu_{R}$~\cite{gell-mann}.

Phenomenologically, the introduction of three families for the right-handed
neutrinos brings a new matrix for the Yukawa couplings in the lepton sector,
similar to the one in the quark sector. A simultaneous diagonalization of both
matrices is quite accidental and, as in the quark sector, the new leptonic 
Yukawa couplings cause lepton-flavour violation. This phenomenon could exist in
a supersymmetric version of the Standard Model with the introduction of 
right-handed neutrinos~\cite{pilaftsis}. It can also occur naturally 
in one of the $SUSY-GUT$ models. The simplest extension of the Minimal 
Supersymmetric Standard Model which would include right-handed neurinos 
naturally is the one based on the left-right symmetric extension of the Standard Model.
We will investigate here lepton-flavor violating decays of the 
$Z$-boson in a fully left-right  supersymmetric  model .

The Left-Right Supersymmetric Model $(LRSUSY)$ is an extension of the Minimal 
Supersymmetric Standard Model  based on left-right symmetry.
$LRSUSY$ shares some of the attractive  properties of 
the $MSSM$ , like providing a natural solution for the gauge hierarchy
problem; and, in addition, $LRSUSY$ supresses naturally rapid proton decay,
by disallowing terms in the Lagrangian that explicitly violate either 
baryon or lepton numbers~\cite{frank1}.It gauges the only quantum 
number left ungauged, $B-L$. The $LRSUSY$ model shares some of 
the attractive 
features of the original left-right symmetric model~\cite{mohapatra},
such as providing a possible explanation
for the strength of $CP$ violation. It could  be viewed as a low-energy 
realization of certain $SUSY-GUTs$, such as $SO(10)$ or $SU(5)$.
So far there is no 
experimental evidence for the right-handed interactions predicted by the 
$SU(2)_{L}\times SU(2)_{R}\times U(1)_{B-L}$ theory, let alone by supersymmetry.
Yet the foundation of $LRSUSY$ has so many attractive features that the model 
deserves some experimental and theoretical investigation. The next generation 
of linear colliders will provide an excellent opportunity for such a study. 
The theoretical and experimental challenge lies in finding distinctive features
for the left-right supersymmetric model, which allow it to be distinguished 
from both the SUSY version of the Standard Model and from the non-supersymmetric
version of the left-right theory~\cite{frank3}. Lepton-flavour violation decays are just the right type of such phenomena.
The  LRSUSY model provides a natural framework for large lepton flavor-violating
effects through two mechanisms: on one hand it can give rise to 
a leptonic decay width of the $Z$-boson through both left-handed and 
right-handed scalar lepton mixing , on the other hand it contains 
lepton-flavor-blind higgsinos which couple to leptons only and enhance 
lepton-flavor violation.

Our paper is organized as follows: in Section 2, we  describe the $LRSUSY$
model; in Section 3 we  discuss the supersymmetric contributions (including 
both the $MSSM$ and the $LRSUSY$ contributions) to
the decay rate $Z \rightarrow l_{1} \bar l_{2}$. We will conclude in Section 4 with 
numerical estimates and discussion. 

\section{\bf The Left-Right Supersymmetric Model}

The $LRSUSY$ model, based on $SU(2)_{L}\times SU(2)_{R}\times U(1)_{B-L}$,
has matter 
doublets for both left- and right- handed fermions and the corresponding left- 
and right-handed scalar partners (sleptons and squarks)~\cite{frank3}.
In the gauge sector, 
corresponding to $SU(2)_{L}$ and $SU(2)_{R}$, there are triplet
gauge bosons $(W^{+,-},W^{0})_{L}$, $(W^{+,-},W^{0})_{R}$ and a singlet gauge
boson $V$ corresponding to $U(1)_{B-L}$, together with their superpartners. 
The Higgs sector of this model 
consists of two Higgs bi-doublets, $\Phi_{u}(\frac{1}{2},\frac{1}{2},0)$ and 
$\Phi_{d}(\frac{1}{2},\frac{1}{2},0)$, which are required to give masses to 
both the up and down quarks. In addition, the spontaneous symmetry breaking of
the group 
$SU(2)_{R}\times U(1)_{B-L}$ to the hypercharge symmetry group $U(1)_{Y}$ is 
accomplished by introducing the Higgs triplet fields $\Delta_{L}(1,0,2)$ 
and $\Delta_{R}(0,1,2)$. The choice of the triplets (versus four doublets)
is preferred because with this choice a large Majorana mass can be generated
for the right-handed neutrino and a small one for the left-handed neutrino~
\cite{mohapatra}.
In addition to the triplets $\Delta_{L,R}$, the model must contain two 
additional triplets $\delta_{L}(1,0,-2)$ and $\delta_{R}(0,1,-2)$ , with 
quantum number $B-L= -2$ to insure cancellation of the anomalies that would 
otherwise occur in the fermionic sector. Given their strange quantum numbers,
the $\delta_{L}$ and $\delta_{R}$ do not couple to any of the particles in the 
theory so their contribution is negligible for any phenomenological studies.

As in the standard model, in order to preserve $U(1)_{EM}$ gauge 
invariance,  only the neutral Higgs fields aquire non-zero vacuum 
expectation values $(VEV's)$. These values are:
\begin{eqnarray}
< \Delta_{L}> = \left(\begin{array}{cc}
0&0\\v_{L}&0
\end{array}\right),
<\Delta_{R}> = \left (\begin{array}{cc}
0&0\\v_{R}&0
\end{array}\right)~\rm{and}~
< \Phi> = \left (\begin{array}{cc}
\kappa&0\\0&\kappa' e^{i\omega}
\end{array}\right)
\nonumber
\end{eqnarray}
$<\Phi>$ causes the mixing of $W_{L}$ and $W_{R}$ bosons with $CP$-violating 
phase $\omega$. In order to simplify, we will take the $VEV's$ of the Higgs fields as: $< \Delta_{L}> = 0$ and 
\begin{eqnarray}
<\Delta_{R}> = \left (\begin{array}{cc}
0&0\\v_{R}&0
\end{array}\right),
< \Phi_{u}> = \left (\begin{array}{cc}
\kappa_{u}&0\\0&0
\end{array}\right)~\rm{and}~
< \Phi_{d}> = \left (\begin{array}{cc}
0&0\\0&\kappa_{d}
\end{array}\right)
\nonumber
\end{eqnarray}
Choosing $v_{L} =\kappa' =0$ satisfies the more loosely required hierarchy
$v_{R}~\gg~max(\kappa,\kappa')~\gg~v_{L}$ and also the required cancellation 
of flavor-changing neutral currents. The Higgs fields aquire non-zero $VEV's$ 
to break both parity and $SU(2)_{R}$.
In the first stage of breaking the right-handed gauge bosons $W_{R}$ and $Z_{R}$
aquire masses proportional to $v_{R}$ and become much heavier than the usual
(left-handed) neutral gauge bosons $W_{L}$ and $Z_{L}$, which pick up masses 
proportional to $\kappa_{u}$ and $\kappa_{d}$ at the second stage of breaking.
~\cite{frank1}

The supersymmetric sector of the model, 
while preserving left-right symmetry, has four singly-charged
charginos ( corresponding to $\tilde\lambda_{L}, 
\tilde\lambda_{R}, \tilde\phi_{u}$, and
$\tilde\phi_{d}$), in addition to $\tilde\Delta_{L}^-$ , 
$\tilde\Delta_{R}^-$ , $\tilde\delta_{L}^-$ and $\tilde\delta_{R}^-$.
The model also has eleven neutralinos, corresponding to $\tilde\lambda_{Z}$, 
$\tilde\lambda_{Z\prime}$,
$\tilde\lambda_{V}$ ,  $\tilde\phi_{1u}^0$ ,$\tilde\phi_{2u}^0$ ,
$\tilde\phi_{1d}^0$ , $\tilde\phi_{2d}^0$, $\tilde\Delta_{L}^0$,
$\tilde\Delta_{R}^0$ $\tilde\delta_{L}^0$, and
$\tilde\delta_{R}^0$. It has been shown that in the scalar sector, 
the left-triplet 
$\Delta_{L}$ couplings can be neglected in phenomenological analyses of 
muon and tau decays~\cite{pilaftsis}. Also $\Delta_{L}$ 
is not necessary for symmetry breaking~\cite{huitu}; 
it is introduced only for preserving left-right symmetry.
We will therefore neglect the couplings of $\Delta_{L}$ 
in the fermionic sector.

The doubly charged $\Delta_{R}^{--}$ is however very important: it carries 
quantum number $B-L $ of 2 and  couples only to leptons, therefore
breaking lepton-quark universality. It and its supersymmetric partner could,
as will be seen in the next 
section, play an important role in flavour-violating leptonic decays.

In the scalar matter sector, the $LRSUSY$ contains two left-handed and two 
right-handed scalar fermions as partners of the ordinary leptons and quarks, 
which themselves
come in left- and right-handed doublets. In general the left- and right-handed  
scalar leptons will mix together. Some of the effects of this mixings, such as 
enhancement of the anomalous magnetic moment of the muon, have been discussed 
elsewhere~\cite{frank1}. Only global lepton-family-number violation
would prevent $\tilde{e}$
, $\tilde{\mu}$ and $\tilde{\tau}$ to mix arbitrarily. Permitting this mixing
to occur, we could expect small effects to occur in the non-supersymmetric 
sector, such as radiative muon or tau decays, in addition to other nonstandard 
effects such as massive neutrino oscillations and violation of lepton number 
itself. But, in general, allowing for the mixing, 
we have six charged-scalar lepton states (involving 15 real angles 
and 10 complex phases) and six scalar neutrinos (also involving 15 real 
angles and 10 complex phases).
In order to reduce the (large) number of parameters
we shall assume in what folows that only two generations of
scalar leptons (the heaviest)
mix significantly~\cite{levine} . The mixings are as follows:
$\tilde\mu_{L,R}$ and $\tilde\tau_{L,R}$ with angle $\theta_{L,R}$;
$\tilde\nu_{\mu_{L,R}}$ and $\tilde\nu_{\tau_{L,R}}$ with angle $\alpha_{L,R}$;
so that, for example:
$$\tilde l_{L_{1}} = \tilde\mu_{L}cos\theta_{L} + 
\tilde\tau_{L}sin\theta_{L}$$
$$\tilde l_{L_{2}} = - \tilde\mu_{L}sin\theta_{L} + 
\tilde\tau_{L}cos\theta_{L}$$ 
and similarly for $\tilde l_{R_{1,2}}$ and $\tilde\mu_{L_{1,2}}$ and
$\tilde\mu_{R_{1,2}}$. These states are the physical mass eigenstates.

Next we consider the implications of the above-mixing in the $LRSUSY$ in
lepton-flavor violating decays of the $Z_{L}$ boson.

\section{\bf Lepton Number Violating Decays}

There are  two types of contributions to the lepton-number violating
decays: one coming from the non-SUSY sector of the left-right model, the other
type coming from the SUSY sector, through the contributions of supersymmetric 
partners. To keep the parameters to a minimum, we choose to 
evaluate analytically the contribution to amplitude for the 
decay $Z \rightarrow l_{1}\bar{l}_{2}$ due to the mixing of scalar leptons and scalar
neutrinos and due 
to the charginos and neutralinos in the model. We note that the mass parameters 
which determine the contribution in the SUSY sector are independent ( in broken 
supersymmetry) of the ones in the non-SUSY sector. For completeness, we refer 
the reader to~\cite{pilaftsis}.

In general , the branching ratio for this process can be written as:
$$B(Z \rightarrow \bar{l_{1}}l_{2} + l_{1}\bar{l_{2}}) =\frac{\alpha_{w}^3}
{48\pi cos_{w}^2}\frac{M_{Z}}{\Gamma_{Z}}[|{\Gamma_{l_{1}l_{2}}^L|}^2 + 
|{\Gamma_{l_{1}l_{2}}^R|}^2],$$
where the nonoblique functions 
$\Gamma_{l_{1}l_{2}}^L$ and $ \Gamma_{l_{1}l_{2}}^R$ depend on the $V-A$ or
$V+A$ character of the theory. Such a branching ratio is restricted by $LEP$
to be \it{e.g.}\rm~ $B(Z \rightarrow e\tau) \le 10^{-5}$~\cite{particle data}. 

We assume that the leptonic masses are small (and therefore negligible) compared
to at least one mass of a supersymmetric particle in the process.
 
There are 9 relevant diagrams that contribute to 
$\Gamma_{l_{1}\bar{l}_{2}}^L$ and $ \Gamma_{l_{1}\bar{l}_{2}}^R$
and they are listed in Fig.~1. 
The contributions from the individual graphs are listed below. First , the 
left-handed contributions which are similar, although not identical, to the ones
in the $MSSM$~\cite{levine} :

\begin{eqnarray}
\Gamma_{l_{1}\bar{l}_{2}}^{L}(A)&=&\frac{-1}{2}\frac{ie^3}{32\pi^2\sin^2
\theta_{w} \sin 2\theta_{w}}
\sin 2\theta_{L}[\sum_{i=1}^4(|\tan\theta_{w}N_{i1}+N_{i2}|^2)\cos 
2\theta_{w}
\tilde\lambda_{l_{L_{1}}} \nonumber\\
& &C_{24}(\tilde\lambda_{i}^0,
\tilde\lambda_{l_{L_{1}}},\tilde
\lambda_{l_{L_{1}}})]
\end{eqnarray}
\begin{eqnarray}
\Gamma_{l_{1}\bar{l}_{2}}^{L}(B) \hspace{0.2cm} = \hspace{0.2cm} \frac{ie^3}{32\pi^2\sin^2\theta_{w}
\sin2\theta_{w}}
\sin 2\alpha_{L}\sum_{i=1}^4(|V_{i1}|^2)\tilde\lambda_{\nu_{L_{1}}}
C_{24}(\tilde\lambda_{i}^{-},
\tilde\lambda_{\nu_{L_{1}}},\tilde
\lambda{\nu_{L_{1}}})
\hspace{0.3cm}
\end{eqnarray}
\begin{eqnarray}
\Gamma_{l_{1}\bar{l}_{2}}^{L}(D)&=&\frac{ie^3}{32\pi^2\sin^2\theta_{w}
\sin2\theta_{w}}
\sin2\theta_{L} \hspace{7cm}
\nonumber\\
& & [\sum_{i=1}^4\sum_{j=1}^4(\tan\theta_{w}N_{i1}^*+N_{i2}^*)
(\tan\theta_{w}N_{j1}+N_{j2}) 
\nonumber\\
&  & O_{ij}^{L''}(2C_{24}(\tilde\lambda_{l_{L_{1}}},
\tilde\lambda_{i}^0, \tilde\lambda_{j}^0) -1/2+\lambda_{z}(C_{23}
(\tilde\lambda_{l_{L_{1}}},\tilde\lambda_{i}^0,\tilde\lambda_{j}^0)
\nonumber\\ 
&  & -C_{22}(\tilde\lambda_{l_{L_{1}}},\tilde\lambda_{i}^0,\tilde\lambda_{j}^0)) - O_{ij}^{R''} \sqrt{\tilde\lambda_{i}^0\tilde\lambda_{j}^0}C_{0}
(\tilde\lambda_{l_{L_{1}}},\tilde\lambda_{i}^0,\tilde\lambda_{j}^0)]
\end{eqnarray} 
\begin{eqnarray}
\Gamma_{l_{1}\bar{l}_{2}}^{L}(E)& = &\frac{2ie^3}{32\pi^2\sin^2\theta_{w}\sin2
\theta_{w}}
\sin2\alpha_{L}[\sum_{i=1}^4\sum_{j=1}^4(V_{i1}^*V_{j1}) \hspace{4cm}
\nonumber\\
&  &O_{ij}^{L'}(2C_{24}(\tilde\lambda_{\nu_{L_{1}}},
\tilde\lambda_{i}^-,\tilde\lambda_{j}^+) -1/2+\lambda_{z}(C_{23}
(\tilde\lambda_{\nu_{L_{1}}},\tilde\lambda_{i}^-,\tilde\lambda_{j}^+)
\nonumber\\
&  &  - C_{22}
(\tilde\lambda_{\nu_{L_{1}}},\tilde\lambda_{i}^-,\tilde\lambda_{j}^+)) -
O_{ij}^{R'}
\sqrt{\tilde\lambda_{i}^-\tilde\lambda_{j}^+}C_{0}
(\tilde\lambda_{\nu_{L_{1}}},\tilde\lambda_{i}^-,\tilde\lambda_{j}^+)]
\hspace{2.7cm} 
\end{eqnarray}

\begin{eqnarray}
\Gamma_{l_{1}\bar{l}_{2}}^{L}(G)& = &\frac{1}{2}\frac{ie^3}{32\pi^2\sin^2
\theta_{w}\sin2\theta_{w}}
\sin2\theta_{L}\sum_{i=1}^4(|\tan\theta_{w}N_{i1}+N_{i2}|^2)\cos2\theta_{w}
\nonumber\\ 
&  & 
B_{1}(0,\tilde\lambda_{i}^0,
\tilde\lambda_{l_{L_{1}}})
\end{eqnarray}

\begin{eqnarray}
\Gamma_{l_{1}\bar{l}_{2}}^{L}(H) \hspace{0.2cm}=\hspace{0.2cm} \frac{1}{2}\frac{ie^3}{32\pi^2\sin^2
\theta_{w}\sin2\theta_{w}}
\sin2\alpha_{L}\sum_{i=1}^4 \tilde\lambda_{i}^-(|V_{i1}|^2) 
 B_{1}(0,\tilde\lambda_{i}^{-},
\tilde\lambda_{\nu_{R_{1}}})
\end{eqnarray}

where $\lambda_{n}= m_{n}^2/M_{W_{L}}^2$, the functions $C_{ij}
(\lambda_{i},\lambda_{j},\lambda_{k})$
are the three-point functions , and $B_{1}(0,\lambda_{i},\lambda_{j})$ is a 
two-point function associated with the self-energy graphs. We follow the notation and conventions of Ref.~\cite{kniehl}, to which we refer the reader for further details.
In $LRSUSY$ mixing of right-handed scalar leptons as well as gauginos
induce a nonuniversal coupling of the $Z_{L}$ , $\Gamma^R$ as shown
in Fig.~1. The contributions are as given below:
\begin{eqnarray}
\Gamma_{l_{1}\bar{l}_{2}}^{R}(A) & = & 2\frac{ie^3}{32\pi^2\sin^2\theta_{w}
\sin2\theta_{w}}
\sin2\theta_{R}\sum_{i=1}^4(|\tan\theta_{w}N_{i1}|^2)2sin^2\theta_{w}
\tilde\lambda_{l_{R_{1}}} \hspace{1cm}
\nonumber\\
&  & C_{24}(\tilde\lambda_{i}^0,
\tilde\lambda_{l_{R_{1}}},\tilde
\lambda_{l_{R_{1}}})
\end{eqnarray}
\begin{equation}
\Gamma_{l_{1}\bar{l}_{2}}^{R}(B) \hspace{0.2cm}=\hspace{0.2cm} \frac{ie^3}{32\pi^2\sin^2\theta_{w}
\sin2\theta_{w}}
\sin2\alpha_{R}\sum_{i=1}^4(|V_{i2}|^2)
\tilde\lambda_{\nu_{R_{1}}} C_{24}(\tilde\lambda_{i}^{-},
\tilde\lambda_{\nu_{R_{1}}},\tilde
\lambda_{\nu_{R_{1}}}) \hspace{0.7cm}
\end{equation}
\begin{equation}
 \Gamma_{l_{1}\bar{l}_{2}}^{R}(C) \hspace{0.2cm}=\hspace{0.2cm} \frac{ie^3}{32\pi^2\sin^2\theta_{w}
\sin2\theta_{w}}
\sin2\theta_{R}\tan^2\theta_{k}\sin^2\theta_{w}
\tilde\lambda_{l_{R_{1}}} C_{24}(\tilde\lambda_{\delta}^{--},
\tilde\lambda_{l_{R_{1}}},\tilde
\lambda_{l_{R_{1}}})
\end{equation}
\begin{eqnarray}
\Gamma_{l_{1}\bar{l}_{2}}^{R}(D)& = &\frac{ie^3}{32\pi^2\sin^2\theta_{w}
\sin2\theta_{w}}
\sin2\theta_{L}[\sum_{i=1}^4\sum_{j=1}^4(2\tan\theta_{w}N_{i1}^*)
(2\tan\theta_{w}N_{j1})O_{ij}^{R"}\nonumber\\
& & (2C_{24}(\tilde\lambda_{l_{R_{1}}},
\tilde\lambda_{i}^0,\tilde\lambda_{j}^0) -1/2+\lambda_{z}(C_{23}
(\tilde\lambda_{l_{R_{1}}},\tilde\lambda_{i}^0,\tilde\lambda_{j}^0) 
\nonumber\\
&  &- C_{22}
(\tilde\lambda_{l_{R_{1}}},\tilde\lambda_{i}^0,\tilde\lambda_{j}^0))  
- O_{ij}^{L"}\sqrt{\tilde\lambda_{i}^0\tilde\lambda_{j}^0}C_{0}
(\tilde\lambda_{l_{R_{1}}},\tilde\lambda_{i}^0,\tilde\lambda_{j}^0)]
\end{eqnarray} 
\begin{eqnarray}
\Gamma_{l_{1}\bar{l}_{2}}^{R}(E) & = &\frac{2ie^3}{32\pi^2\sin^2\theta_{w}
\sin2\theta_{w}}
\sin2\alpha_{R}[\sum_{i=1}^4\sum_{j=1}^4(V_{i2}^*V_{j2})O_{ij}^{R'}
\hspace{3cm}
\nonumber\\
&  & (2C_{24}(\tilde\lambda_{\nu_{R_{1}}},
\tilde\lambda_{i}^-,\tilde\lambda_{j}^+) -1/2+\lambda_{z}(C_{23}
(\tilde\lambda_{\nu_{R_{1}}},\tilde\lambda_{i}^-,\tilde\lambda_{j}^+) 
\nonumber\\
&  & - C_{22}(\tilde\lambda_{\nu_{R_{1}}},\tilde\lambda_{i}^-,
\tilde\lambda_{j}^+)) 
 - O_{ij}^{L'}
\sqrt{\tilde\lambda_{i}^-\tilde\lambda_{j}^+}C_{0}
(\tilde\lambda_{\nu_{R_{1}}},\tilde\lambda_{i}^-,\tilde\lambda_{j}^+)]
\end{eqnarray}
\begin{eqnarray}   
\Gamma_{l_{1}\bar{l}_{2}}^{R}(F)& = &\frac{1}{2}\frac{ie^3}{32\pi^2\sin^2
\theta_{w}\sin2\theta_{w}}
\sin2\theta_{R}\sin^2\theta_{w}\tan^2\theta_{k}\tilde\lambda_{\delta}
\hspace{2.5cm}
\nonumber\\
&  & [2C_{24}(\tilde\lambda_{\nu_{R_{1}}},
\tilde\lambda_{\delta},\tilde\lambda_{\delta}) -1/2+
\nonumber\\
&  &\lambda_{z}(C_{23}
(\tilde\lambda_{\nu_{R_{1}}},\tilde\lambda_{\delta},\tilde\lambda_{\delta}) 
- C_{22}
(\tilde\lambda_{\nu_{R_{1}}},\tilde\lambda_{\delta},\tilde\lambda_{\delta}))] 
\end{eqnarray}   
\begin{eqnarray}
\Gamma_{l_{1}\bar{l}_{2}}^{R}(G) & = & 2\frac{ie^3}{32\pi^2\sin^2\theta_{w}\sin2
\theta_{w}}
\sin2\theta_{R}\sum_{i=1}^4\tilde\lambda_{i}^0
(|\tan\theta_{w}N_{i1}|^2)2\sin^2\theta_{w}
\hspace{0.5cm}
\nonumber\\ 
&  & B_{1}(0,\tilde\lambda_{i}^0,
\tilde\lambda_{l_{R_{1}}})
\end{eqnarray}
\begin{equation}
\Gamma_{l_{1}\bar{l}_{2}}^{R}(H) \hspace{0.1cm} = \hspace{0.1cm} 4\frac{ie^3}{32\pi^2\sin^2\theta_{w}\sin2
\theta_{w}}
\sin2\alpha_{R}\sum_{i=1}^4 \tilde\lambda_{i}^-(|V_{i2}|^2)
 B_{1}(0,\tilde\lambda_{i}^{-},
\tilde\lambda_{\nu_{R_{1}}})
\end{equation}
\begin{equation}
\Gamma_{l_{1}\bar{l}_{2}}^{R}(I) \hspace{0.2cm} = \hspace{0.2cm} \frac{-1}{4}\frac{ie^3}{32\pi^2\sin^2
\theta_{w}\sin2\theta_{w}}\tan^2\theta_{k}
\sin2\theta_{R}\tilde\lambda_{\delta}
 B_{1}(0,\tilde\lambda_{\delta},
\tilde\lambda_{\delta})
\hspace{0.7cm}
\end{equation}

Here $O_{ij}$ and $N_{ij}$ arise from neutralino mixing and coupling 
at the vertices; $O\prime_{ij}$ and $V_{ij}$ , $U_{ij}$ arise from 
chargino mixing at the vertices. The expressions for these matrix elemens are 
rather lengthy; they have been calculated previously, in a work dealing
in detail with chargino and neutralino masses, and are listed in ~\cite{frank2}.

\section{\bf Numerical Results and Discussion}

As seen in the previous section, the left-right supersymmetric contributions
to the branching ratio $BR(Z \rightarrow l_{1}\bar{l_{2}})$ depend on a plethora of parameters: 
chargino and neutralino masses and mixing matrix elements, as well as
sleptons and sneutrino masses and mixing angles.
An exact evaluation of this cross section would require fixing all the parameters: clearly an 
impossible task. We attempt to estimate the possible values of the branching 
ratio in order to get an idea for the range of parameters and then compare the result 
with current experimental data: for example with the branching ratio 
$BR(Z \rightarrow \tau^{\pm}e^{\mp}) \le 1.3\times 10^{-5}$. Similar results are 
obtained for $BR(Z \rightarrow \mu^{\pm}\tau^{\mp}) \le 1.\times 10^{-5}$ and 
$BR(Z \rightarrow e^{\pm}\tau^{\mp}) \le 6\times 10^{-6}$~\cite{particle data}. 

In particular, we would like to get an estimate for the case in which this 
branching ratio is large, perhaps close to the experimental bound. We shall make
several simpifying assumptions with this in mind.

First, from the expressions for the matrix elements, it can be seen that
the largest cross 
section  occurs when the mass splittings between $\tilde l_{L_{1,2}}$, 
$\tilde l_{R_{1,2}}$ ,  $\tilde \nu_{L_{1,2}}$ and
$\tilde \nu_{R_{1,2}}$ are large. We shall therefore assume that the scalar 
leptons and neutrinos mix maximally, that is,
$\theta_{L} = \theta_{R} =\alpha_{L}= \alpha_{R}= \pi/4$. Also we 
assume maximal mass splitting for the scalar leptons and sneutrinos, that is, 
we assume that 
$\tilde \nu_{L_{1}}$, $\tilde \nu_{R_{1}}$, 
$\tilde l_{L_{1}}$ and  $\tilde l_{R_{1}}$ are relatively light, but  
$\tilde \nu_{L_{2}}$,
$\tilde \nu_{R_{2}}$, $\tilde l_{L_{2}}$ and  $\tilde l_{R_{2}}$ are
very heavy , and therefore decouple.

Second, the cross section will be smallest for the lightest allowable 
supersymmetric partners. As the mass of any particle becomes large, the 
contributions from the subprocesses involving that particle become negligibly 
small.

Since leptonic masses are taken to be small compared to that of 
their supersymmetric partners, the terms proportional to lepton masses arising
 from the higgsino component of a vertex (either chargino or neutralino) can 
be ignored.

We base our estimates for chargino and neutralino masses and mixing matrices 
on previous work done on the subject. For completness, we refer the reader to \cite{frank2}. It suffices to mention here that both 
chargino and neutralino 
masses, as well as their mixing matrices depend on the following 
parameters :$M_{L}$ , 
the left-handed gaugino mass parameter,  
$M_{R}$,the right-handed gaugino mass parameter, 
$\mu$, the higgsino mass parameter, and 
$\tan\theta_{k}= \kappa_{u}/\kappa_{d}$.
In addition the cross section will depend on the mass of the $\Delta_{L}^{--}$
which is otherwise unrestricted.

We shall examine the cross section as a function of all the masses, which is 
more illuminating than a function of $M_{L}$, $M_{R}$, $\mu$ and  
$\tan\theta_{k}$. For simplification we will take $\tan\theta_{k}=2$ in all
our considerations.

To get a feeling for the size of the supersymmetric contributions we first 
choose an  ``unbroken supersymmetry" limit, in which the first two neutralinos 
are degenerate in mass  with the photon and $Z_{L}$ boson, the first two 
charginos are degenerate in mass with the $W_{L}$ boson , all the ``1" scalar 
leptons and scalar neurinos have the same mass , and all the other 
supersymmetric particles are much heavier  and decouple. The only 
supersymmetry breaking terms are explicit mass terms for the scalar 
leptons and neutrinos. The results are shown in Fig.~2. It can bee seen that
for light scalar partners in this idealized scenario,
the branching ratio  $Z \rightarrow l_{1} \bar{l_{2}}$ can reach values
close to the experimental limits. 

After getting a rough estimate for the size of the branching ratio, we analyse
the effects of varying one of the masses while keeping the others 
constant. We do this to investigate the dependence of the branching ratio on
the chargino and  neutralino masses , as well as 
scalar leptons and scalar neutrino masses. We restrict
our analysis to the light mass region of the supersymmetric particles 
in the hope  that effects measured there could be noticeable.

In Fig.~3 we plot the dependence of the branching ratio on the 
chargino and neutralino masses. It is impossible to treat
separately the chargino and neutralino masses, since as explained above,
they depend on the same set of parameters. To simplify, we
consider the influence of the lightest two charginos. Taking
$m_{\tilde\chi_{1}^\pm}$ to be the variable, we see that as expected,
the cross section decreases with increasing  $m_{\tilde\chi_{1}^\pm}$  and could reach
$3\times10^{-5}$ for all masses being at their lowest experimental bound. 
(We could have chosen in much the same way to plot the dependence of the branching ratio 
on the lightest neutralino mass $m_{\tilde\chi_{1}^0}$ ).
In many models the first
neutralino (the photino) is assumed to be the LSP and its mass is restricted to be
$\ge 25~GeV$; such values can be obtained in the $LRSUSY$ model 
and are consistent with Fig.~3.  The cross section again can reach
values close to the experimental bounds.

Fig.~4  shows  the dependence of the cross section on 
the scalar neutrino masses.
( Assuming the left and right scalar neutrinos to be degenerate in mass).
We also take  the left and right  
scalar leptons to be degenerate in mass around $100~GeV$. This is for the  
purpose of increasing the total cross section, since the right-handed sleptons 
contribute only to $\Gamma_{l_{1}\bar{l}_{2}}^R$ and the left-handed ones only to 
$\Gamma_{l_{1}\bar{l}_{2}}^L$. Again , these are idealized assumptions,
the cross 
section varies weakly with these masses, but it can reach values near the 
experimental bounds.

The most interesing dependence is perhaps the one on the mass of the 
doubly-charged $\tilde\Delta_{R}^{--}$ higgsino. This chargino does not mix with other
ones because of its charge, and it can significantly increase the branching 
ratio $BR(Z  \rightarrow l_{1} \bar{l_{2}})$ through  scalar-leptons-scalar-leptons loops,
as well as through the self energy graphs. The dependence of the cross section 
on the mass of the $\tilde\Delta_{R}^{--}$ is shown in Fig.~5. The branching ratio 
can exceed the experimental limits unless the mass of the $\tilde\Delta_{R}^{--}$ 
is larger than 120 GeV. This seems to be the only definite restriction of 
this analysis. It seems to persist even when $\tan\theta_{k}$ is varied(in the region 2 to 5).
When $\tan\theta_{k}$ varies, the chargino and neutralino masses become large and their 
contribution smaller, but this is offset by an increase in the contribution 
from the $\tilde\Delta_{R}^{--}$.

In conclusion , we have shown that the Left-Right Supersymmetric Model, unlike
the $MSSM$ ,  can give significant contributions to lepton-number-violating
decays of the $Z_{L}$-boson, especially for light (near experimental bounds) charginos
and neutralinos.  It seems possible to observe such a decay rate at SLC, 
even if the only contribution to these decays arises from slepton and sneutrino 
mixing. The present limits on the masses and leptonic decays seem to restrict 
the mass of the doubly-charged higgsino to at least 120 GeV.
 
\section{\bf Acknowledgements}
M.F. would like to thank Richard Hall for help with Mathematica and Harith Saif
for past collaborations on this subject, especially on~\cite{frank3}. M. F. would like to thank NSERC for partial financial support. The work of H. H. is supported in part by the United States Department of Energy under grant no.  DE-FG02-84ER40153.

\newpage
\section{\bf List of Figures}
\medskip

{\bf Figure 1} : Feynman graphs contribuing to the effective nonoblique 
$Zl_{1}\bar{l_{2}}$ coupling in $LRSUSY$.
\medskip

\noindent
{\bf Figure 2} : Branching Ratio $B(Z \rightarrow l_{1}\bar{l_{2}})$ in the "supersymmetry"
 limit. Here we vary all slepton and sneutrino masses (assumed to be equal), 
while keeping the chargino 
and neutralino masses at their $SUSY$ limit : $m_{\tilde\chi_{1}^0}$ = 
$m_{\tilde\gamma}$ = 0, $m_{\tilde\chi_{2}^0}$ = 
$m_{\tilde Z_{L}}$ = $m_{Z}$, $m_{\tilde\chi_{1}^\pm}$ = $m_{\tilde\chi_{2}^\pm}$ = 
$m_{\tilde W_{L}}$ = $m_{W_{L}}$,  all the other superpartners decouple.
 As defined in the text, $\lambda_{n}= m_{n}^2/M_{W_{L}}^2$.
\medskip

\noindent
{\bf Figure 3} : Branching Ratio $B(Z \rightarrow l_{1}\bar{l_{2}})$ as a function
of the lightest chargino mass  $m_{\tilde\chi_{1}^\pm}$, showing the
dependence of the cross section on the chargino-neutralino mass parameters.
Here $\lambda_{\tilde\chi_{1}^\pm}= m_{\tilde\chi_{1}^\pm}^2/M_{W_{L}}^2$.
A similar graph could be obtained if one plotted the branching ratio as a 
function of the lightest neutralino mass. Here the scalar partners, the slepton 
and the sneutrino are assumed to have masses close to the experimental bounds:
$m_{\tilde l} = 100~GeV $ and $m_{\tilde\nu} = 45~GeV$. 
\medskip

\noindent
{\bf Figure 4} : Branching Ratio $BR(Z \rightarrow l_{1}\bar{l_{2}})$ as a function
of the sneutrino mass  $m_{\tilde\nu}$. Here all the 
chargino and neutralino have $SUSY$ limit masses: $m_{\tilde\chi_{1}^0}$ = 
$m_{\tilde \gamma}$ = 0, $m_{\tilde \chi_{2}^0}$ = 
$m_{\tilde Z_{L}}$ = $m_{Z}$, $m_{\tilde \chi_{1}^\pm}$ = $m_{\tilde \chi_{2}^\pm}$ 
= $m_{\tilde W_{L}}$ = $m_{W_{L}}$ , all the other superpartners decouple.
The slepton mass is near the lowest experimental bound :
 $m_{\tilde l} = 100~GeV$ .
\medskip

\noindent
{\bf Figure 5} : Branching Ratio $B(Z \rightarrow l_{1}\bar{l_{2}})$ as a function
 of the doubly charged higssino , $\tilde\Delta_{L}^{++}$ mass . Here all the 
chargino and neutralino have $SUSY$ limit masses: $m_{\tilde\chi_{1}^0}$ = 
$m_{\tilde\gamma} = 0$, $m_{\tilde\chi_{2}^0}$ = 
$m_{\tilde Z_{L}} = m_{Z}$, $m_{\tilde \chi_{1}^\pm}$ = $m_{\tilde \chi_{2}^\pm}$ 
= $m_{\tilde W_{L}}$ = $m_{W_{L}}$, all the other superpartners decouple.
Here the slepton 
and the sneutrino are assumed to have masses close to the experimental bounds:
$m_{\tilde l} = 100~GeV$ and $m_{\tilde \nu} = 45~GeV$. 

\end{document}